\documentclass[12pt]{article}
\usepackage{a4}
\usepackage{amssymb}
\begin{document}

\title{Some bounds extracted from a quantum of area}
\author{Alejandro Rivero\thanks{Zaragoza University at Teruel.  
           {\tt arivero@unizar.es}}}

\maketitle

\begin{abstract}
Asking very elementary relativistic quantum mechanics to meet quantum 
of area and time, it is possible to derive at a general level: a) the seesaw 
bound for the mass of neutrinos, and b) the need of a gauge group at 
energies below Planck mass.
\end{abstract}

Having the seesaw mechanism as an explicit realization, a general lower bound on the mass of neutrinos has been traditionally derived from Planck mass (see e.g. R.E. Schrock mini-review on neutrinos for \cite{pdg})  so that $m_\nu \approx a m_{ew}^2/ \bar M_P$, with $m_{ew}$ the electroweak scale, $\bar M_P$ a generic new physics scale whose maximum value is Planck mass, and $a$ a dimensionless constant, probably a Yukawian coupling.  

Now it is obvious that any argument asking an electroweak force $G_F m^2_\nu$ to be of order $a m_\nu/\bar M_P$ will serve to obtain this bound. The later quotient can be argued as a quotient of energies or as a quotient of wavelengths.

It seems interesting to remark that we can derive this condition from the existence of a quantum of area, asking that the measurement of area swept by any orbiting particle must be at least such quantum and assuming this quantum to be directly related to Planck's area. The example reshapes in some way our view of symmetry breaking at the Planck scale, as it happens that below this mass scale gravity becomes too weak to be able to sustain this quantum and it needs to be supplemented by other forces: electromagnetism, strong or electroweak forces. In the case of neutrinos only electroweak interaction is available, and this imposes a bound to neutrino mass. 

Lets think first a classical, non relativistic case. For a force field $F=- K / r^2$ a circular orbit at radius $r$ will have an angular speed $\omega$ such that $K/r^2= m r \omega^2$, and thus it will orbit a length $\delta l =  r  \omega \delta t$ in a time $\delta t$ and to sweep an area $\delta A= \frac 12 r^2 \omega \delta t$. The angular momentum of the particle is $L= m r v = m r^2 \omega$. You can notice that $K= v L$, but we do not use this property in the inequalities.

Now we ask that none of the length dimensions can be smaller than a minimum. Then we have

$$ \frac 12 r^2 \omega \delta t >  l_P^2$$
$$  r  \omega \delta t > l_P$$

$$ \frac 12 {L\over m} v (\delta t)^2 = 
\frac 12 {K\over m} (\delta t)^2 = \frac 12 r^3 \omega^2 (\delta t)^2 > l_P^3 $$

Thus we have a bound on the coupling constant:
$$
K > 2 \, m \, { l_P^3 \over (\delta t)^2}
$$ 

Particularly for a Newtonian interaction $K=GMm$ (and we assume here $M>>m$ or resort to effective mass correction) the bound can be rewritten as a bound on the charge $M$ generating the potential.
$$
 M > 2 {l_P^3 \over G (\delta t)^2}
$$

for small enough $\delta t$. Now the interesting thing is that any upper bound on $\delta t$ will impose a lower bound on $M$ for gravity with a quantum of length to hold. If we use an speed $c_0$ to bound $\delta t < l_P/c_0$ then

$$ 
M > 2 { l_P c_0^2 \over G}  
$$

The point being that for masses under this bound the requirement of a minimal area asks for a supplementary force with coupling $K'$ so that the total coupling $GMm + K'$ will be still strong enough to meet our bound. Of course if $l_P$ is Planck length and $c_0$ is lightspeed then we are telling that any particle with a mass lower than Planck mass {\it must} have extra charges to generate additional fields. At the same time, we are telling that the additional fields (the GUT fields) will start having coupling constant at least of order unit (and as we will discuss, surely no higher) and they will need to keep K at lower energies high enough to meet the bound on $K/m$. 

But in order to do such affirmations we need three things:

- We need to believe in an upper bound for $\delta t$, while we are asking for lower bound in measured geometrical quantities.

- We need to go relativistic to justify the use of $c_0$ at least as an $O(1)$ multiple of $c$. 

- We need to go quantum and use $h$ to justify $l_P$ as coming from the inside of $G$.

Why to bother on it? The possibility of arguing for the origin of GUT forces as gravity breaks below Planck of mass is already enough justification for the effort. But consider again the bound on $K$ and put $m_P={ \hbar \over c l_P}$, then

$$
 K > 2 m l_P c_0^2 = ({ 2 c_0^2 \hbar \over c})  { m \over m_P }   
$$

And think of poor neutral uncoloured neutrino. How can it meet our bound? If we think from Fermi constant 
it generates a Yukawian field force by $Z^0$ exchange at first order... and more complicated SU(2)xU(1) at higher orders, so it is complicated to concrete a value for constant. Using the neutrino mass scale to cancel Fermi scale, we can hand-wave an estimate from

$$K \approx (\hbar c) {m^2 \over m_{ew}^2} $$

but then we have an approximate bound

$$m^2 \gtrsim  m_{ew}^2  { m \over m_P } $$

And thus (I should put a box around it)

$$m_\nu \gtrapprox {m_{ew}^2 \over m_P} $$ 

Of course we have read this goal in the abstract of the paper, so it does not come as a surprise.

As for the three requisites:

Lets dispose of the first requisite by telling that it is a postulate we do not understand. After all, we do not understand neither why a quantum of area is needed, and we started the paper on this way. It could be an operational argument, where area or length are measured outputs and time is required input. 

Only note that if we try to use the velocity of the particle to dispose of $\delta t$ using 
$$v \delta t \approx L_P$$ 
then $v=K/L$ lets us to reverse the role of $K$ in the bound, so we get
$$
L^2 \gtrapprox K {m\over m_P}   
$$ 
which should be valid even for low quantum values of order $h$. Thus we are forced into a boring result: that $K$ should decrease at high energies and while it can increase at low energies, it is not forced to it. Worst, the electroweak neutrino condition would read simply $1 > {m^3 \over m_{EW}^2 m_P}$, a condition compatible with null mass. And $K=GMm$ does not carry us into great deals, it translates to $ L^2 > M m^2 / m_P^3$ so that the requirement of low $L$ states would forbid gravity with masses above the Planck Scale. 

Thus we see that the bound of $\delta t$ should be separated from the kinematics of the particle.

Our second requisite, relativity, is going to be smoother than it could be expected. We had foreseen problems coming from the fact that the coupling constant K has units of angular momentum times speed. Using $c$, it defines a minimum angular moment $L_0=K/c$ so that any orbit having a smaller $L$ will fall into the center, as explained almost a century ago by Darwin and Sommerfeld, and more recently in \cite{basque, tboyer}. Fortunately, this does not imply we haven't got circular orbits, it is only a funny effect of relativistic mass. 


Let be $\gamma={1 \over \sqrt {1 - {v^2/c^2} }}$ and keep the notation $m$ for the rest mass. Now the equilibrium law is $K/r^2= \gamma m r \omega^2$ and our bound is
$$
K > 2 {m \over \sqrt {1 - {v^2/c^2} }} l_P^3 / (\delta t)^2 
= 2 c^2 {m \over \sqrt {1 - {v^2/c^2} }}  l_P 
$$ 
depending on the relativistic mass of the orbiting part. This could endanger our neutrino bound because we are interested on rest mass. Putting $K \approx (m/M_{ew})^2$ we get a $\gamma$ factor hanging:
$$
  m \; m_P >\approx \gamma \; m^2_{EW}
$$
We can save the day arguing that neutrinos will be for sure relativistic particles, so $\gamma \gg 1$ and our bound only risk to become so good that it is experimentally disproved. On the contrary opinion, we can be reminded that the the speed $v$ is going to be $K/L$ and than the coupling is already very small, so that relativistic speeds ask for very very small angular momentum relative to $\hbar$ and it could harbors problems when quantizing.

A completely different focus is to load the charge again on $\delta t$. Remember we started from $\frac 12 r^2 \omega \delta t >  l_P^2$, $  r  \omega \delta t > l_P$. Transforming the second equation to
$$r \omega \gamma \delta t > l_P$$
will cancel the $\gamma$ factor and bring us back to the same formulation that in classical non-relativistic theory.

Our third, and last requisite, is to enter $\hbar$ in play, thus moving to (relativistic) quantum mechanics.

{\bf Digression}

Speaking of the use of $\hbar$, let me to stop to annotate an amateur history. A couple years ago an amateur, Hans de Vries, did the following about the relativistic circular orbit problem \footnote{see http://www.physicsforums.com/showpost.php?p=382642\&postcount=44, 
http://www.physicsforums.com/showpost.php?p=367406\&postcount=36 and 
http://www.physicsforums.com/showpost.php?p=950989\&postcount=197}.

1) Ask the angular momentum to have a very specific value
$$L_s= \sqrt{s (s+1)} \hbar $$

2) Impose the requisite that the rotation frequency is to be equal to the Planckian frequency for the mass of the rotating particle, this is
$$ {v \over r} = {m c^2 \over \hbar} $$ 

3) Note that this requisite implies
$$ L= m \gamma r v^2 = \gamma {v^2 \over c^2} \hbar $$
and use it to solve $v_s$ from the value in step 1)

4) Optionally, calculate $K_s= L_s v_s$, the coupling constant needed to grant the existence of a coulombian circular orbit satisfying 1 and 2. 

5) Take the values $s=1$ and $s=1/2$ and calculate the very funny \footnote{If you are missing the joke, divide the masses of W and Z particles, or check \cite{pdg} for theoretical predictions of the 
Weinberg angle as $s_W^2$ and $\hat s_W^2$} value
$$
 1- ({v_{1/2}\over v_1})^2 = 0.2231013223008663454...
$$ 
It is interesting to note that $K_{1/2}$ is less than $\hbar c$ but $K_1$ is slightly greater than $\hbar c$. 

{\bf End of Digression}

Our argument is based on Keplerian orbits, thus it should make use of Bohr-Sommerfeld relativistic quantum mechanics ("old QM"), a technique good enough to calculate the spectrum of Hidrogen Atom up to fine structure and even to model spin \footnote{L'eon BRILLOUIN, {L'Atome de Bohr}, PUF 1931} but it was not able to pinpoint exactly the absolute vacuum energy, for instance in the harmonic oscillator. It is said (ref. "Sommerfeld Puzzle"?) that actually both problems --neglect of spin and neglect of vacuum $\hbar /2$-- happened to cancel in a hit of Eratosthenian luck, but I am not sure of the details involved. After the Heisenberg-Schroedinger revolution (new QM), B-S model was replaced by Dirac theory, where spin was not modeled ad-hoc but predicted, and then by QFT in order to account for Lamb shift and anomalous magnetic moments, beyond the reach of plain quantum mechanics\footnote{Some coincidence or tuning makes anomalous magnetic moment to look very much as the effective mass correction to a central problem  but that is another history \cite{hans}}. Nowadays it is used only in mathematical physics, under the new disguise of integer cohomology classes on phase space. Readers interested on this formulation (first Chern class) can follow a suggestion of J Baez \footnote{Online, Week 25} who invites to start from the works of B. Konstant \cite{konstant} on Geometric Quantization. 

On the other hand, it should be possible to recast our argument on terms of quantum field theory, perhaps following the lines sketched by Pauli in the letter here in the first Appendix.

In any case, the point is that quantum relativistic Keplerian orbits are described in part II.1 and ff of Sommerfeld work\footnote{A. SOMMERFELD, {\it Zur Quantentheorie der Spektrallinien}, Annalen der Physik (IV Folge), band 51, pp 1-167  }. There we are reminded that angular momentum preservation stills hold \footnote{He uses notation
$$ m r^2 \dot \phi = p,  m= {m_0 \over \sqrt {1 - \beta^2}}, \beta =v/c $$ with an unfortunate $p$ for the angular momentum; we will keep it here but not the rest of the notation}

For a Coulombian central force $F = - K/r^2$ relativity gives us the aforementioned limit angular momentum  $L_{min}\equiv p_0= K/c$, and the quantized angular momentum $p_n={ n h \over 2 \pi}$ is usually expressed as a multiple of this quantity via the famous constant $\alpha \equiv p_0/p_1 = K/\hbar c$. Actually this is the first definition of $\alpha$. When $\alpha > 1$ the Bohr-Sommerfeld theory has not elliptical solutions because the spiraling trajectories of classical relativistic theory are still here and B-S quantization imposes angular momentum to be an exact multiple of reduced Planck constant starting from unity.

So the only problem that quantum theory bring us is an extra requisite
$$\hbar c > K$$
for the theory to be well defined.

Note that in the case of Newtonian gravity with $K= G m_1 m_2$ the combination of bounds closes parameter space: this quantization does not have closed solutions above the Planck scale. 



{\bf Remark 1.} It could be a deeper insight to try to invert the argumentation in order to deduce a minimum of area from the fact of having a non-null neutrino mass \footnote{I thank Sabine H by this suggestion, as well as for generic discussion}


{\bf Remark 2.}
The view of gravity as restored at Planck scale, broken and supplemented at low scale, seems in contradiction with the desire of new physics to appear at Planck scale. But lets note that string theory, that is supposed to be a theory of the new physics at this scale, implies exact General Relativity, so it is more accurate to say that gravity is broken below Planck mass and restored beyond. The same happens with supergravity, which holds exactly only at energies beyond GUT, and is broken at low energy. The paradox seems to be the existence of two infrared limits: the plain need of going to low energy with a single particle, and the need of going to large distances / low energy in the classical limit, where $h\to 0$ asks for a statistic of multiple particles. With a single particle, Lagrangian terms proportional to powers of $h$ should disappear.

{\bf Remark 3}, {\it or digression 2}. We can wonder why are we pivoting around Coulomb potential, in fact other references (eg \cite{basque}) took pains to work with more generality (think if you want to go extra dimensions, for instance). The point is that force, in natural units, has dimensions of $[L]^{-2}$, so when the only source of scale available is the radial distance it is forceful to use inverse square kind of forces. If we want other power, we must provide scale either by using some dimensionful coupling constant (ej put $f=\alpha M/c$ or $f=\alpha /M x^3$) or by taking the scale source from a pair of interacting masses (say $f=\alpha m m'$, $f=\alpha/ m m' x^4$). Of course additional scales can enter with other roles, such as Yukawian cutoffs or approximations to it, wells, etc. No doubt the analysis here can be repeated for each case, but it is not worthwhile without further clarification of the methodology, as seen from the doubts raised along the text.

\appendix

\section*{APPENDIX} 

\section{A letter}

This letter was typewritten the 29th December 1947, and it is cataloged as  [925] PAULI AN L. DE BROGLIE in Pauli's collected letters. I haven't got the opportunity (yet?) to read the papers from De Broglie and Bohr here mentioned. At the time of writing Pauli was excited about the news of the measurement of g-2 (0.00244, by Kusch and Foley) so it is understandable  De Broglie suggestion did not got attention enough. De Broglie's view of quantum mechanics\footnote{L. De BROGLIE, {\it Sur les 'equations et les conceptions g'en'erales de la m'ecanique}, Bull SMF v 58 n 2 (1930) p 1-28 (1930). It is also interesting to check the presentations in the Acad'emie des Sciences, particularly Seance 10 Sept 1923, p 507, where Einstein-Bohr conditions are derived}, close to Schroedinger's, were based on ondulatory optics, the limit $h\to 0$ being analogous to the limit of geometric optics. But in the last years he evolved to support alternate, pretty non-Copenhagian, views and it had could affect the general impact of his work. 

\begin{quotation}

My dear Colleague!

I thank you very much for your manuscript "Sur la compl\'ementarit\'e des id\'ees d'individu et de syst\`eme", which I shall be glad to have in the issue of the "Dialectica" dedicated to the problems of complementarity.

The physical problem, discussed in your paper namely the limitations of the applicability of the concept of constituents due to the condition, that the interaction energy of the constituents of a compound system has to be small in comparison with its restmass* -- can also be discussed from the standpoint of the concept of a field rather than from the standpoint of a potential energy depending on the positions of the constituents. While the latter standpoint is adapted to the situation in non relativistic point mechanics, the former is more conform to the spirit of special relativity. Your condition reappears then in the new form that the reactive force due to the proper field of a constituent particle has to be small in comparison with the force due to the effect of the external field on this particle (If this condition is not fulfilled anymore, the concept of a "constituent particle" certainly ceases to be applicable.)

If we consider the electromagnetic interaction between particles with charge $e$ and restmass $m$, this condition for the external force $K$ becomes essentially with $d=e^2/mc^3$ (c=velocity of light). (N.B. One can be in doubt, whether $d$ should be replaced by the Compton-wave length $t/mc$, but the classical radius $d$ seems to me a more fundamental quantity for this limitation in question.) The new application of the concept of "complementarity" to which this form of the argument is tending, is the mutually exclusive character of the use of the concepts, "field generated by a particle" on the one side, and "charge of the particle" on the other side. It seems that this complementarity is not sufficiently emphasized in the present theory.**

We are also expecting a contribution of Bohr to the issue of the "Dialectica" in question, which, however -- as every redactor knows -- may cause a delay in the appearance of the issue.

The general review of the redaction I shall write myself as soon as all contributions to the issue will have arrived. I am glad, that your contribution is the first one, and I am thanking you again very much for it.

With best wishes for the new year,

sincerely yours

[W. Pauli]

{\small
* In this connection it may be of some interest to point as an example to the singularity in the origin of the wavefunction in the ground state of hydrogen like atoms, which according to Dirac's theory occurs for theoretical value of the nuclear charge Z=137, for which the energy value of this state tends to zero.

** In the summary of the lectures given by Bohr in Cambridge 1946, which is in print, but did not yet appear, he stresses also the dualistic character of the particle and field concepts, which has its roots in the circumstance, that the properties of the particles, like their mass and charge, are defined by the fields of force they produce, or the effects of the fields upon them and, inversely, the fields are themselves only defined through their action on the particles.

}

\end{quotation}

\section{Other old references}

The use of a fundamental length scale was an obvious question since the awareness about a natural system of units by Planck and it has meandered along the history of quantum mechanics. 

In the fifties B. T. Darling did a series of publications on the Physical Review; the first one, Phys Rev 80 p 460-466, quotes a handful of related work by Heisenberg, Schild, Snyder and a lot others, done during the previous decade.

Square root of Newton Constant, Planck length, has been invoked as early as 1937, according \cite{kragh}, to try to interplay between the mass of electron, the one of the muon, and its decay rate... that in turn depend of muon mass and of the Fermi scale. It is not a seesaw-like formula because the powers of mass involved are higher.

In "Poincare and the Quantum Theory", Russel Mc Cormmach reports about Poincare work on a proposal from Nerst trying to obtain Quantum Mechanics as a kind of, say, doubly special relativity with a new mass dilatation formula 
$$m { a \over v \rho} \over \log (1 +  { a \over v \rho} ) $$
suggested in some congress in 1911\footnote{{\it La th\'eorie du rayonnement et les quanta}, ed P. Langevin and M. de Broglie (Paris: Gauthier, 1912), pp 452-453} (Solvay?). It seems that Poincare examined this alternative and concluded it was not equivalent to quantum mechanics. So it could be considered a different kind of quantum of length, perhaps the earliest attempt.

\end{document}